\documentclass[%
 reprint,
showpacs,preprintnumbers,
 amsmath,amssymb,
 aps,
]{revtex4-1}

\usepackage{graphicx}% Include figure files
\usepackage{dcolumn}% Align table columns on decimal point
\usepackage{bm}% bold math
\usepackage{color}

\usepackage{xcolor}
\usepackage{color}
\usepackage{changes}
\begin{document}

\title{Superconducting proximity effect in flat band systems}% Force line breaks with \\
\author{Somayeh Ahmadkhani}
\author{Mir Vahid Hosseini}
 \email[Corresponding author: ]{mv.hosseini@znu.ac.ir}
\affiliation{Department of Physics, Faculty of Science, University of Zanjan, Zanjan 45371-38791, Iran}

%\deleted{proximity}
\begin{abstract}
We study theoretically proximity-induced superconductivity and its inverse effect in dice lattice flat band model by considering Josephson junction with an s-wave pairing in the superconducting leads. Using self-consistent tight-binding Bogoliubov-de Gennes method, we show that there is a critical value for chemical potential of the superconductors depending on paring interaction strength over which for undoped normal region the proximity effect is enhanced. Whereas if the superconductor chemical potential is less than the critical one the proximity effect decreases regardless of normal region doping and in the meanwhile, the pairing amplitude of superconducting region increases significantly. Furthermore, we unveil that the supercurrent passing through the junction is large (vanishingly small) when the superconductor chemical potential is smaller (larger) than the critical value which increases as a function of normal region chemical potential.
\end{abstract}

%\pacs{ }
\maketitle

%%%%%%%%%%%%%%%%%%%%%%%%%%%%%%%%%%%%%%%%%%%%%%%%%%%%%%%%%%%%%%%%%%%%%%%%%%%
\section {Introduction} \label{s1}
%%%%%%%%%%%%%%%%%%%%%%%%%%%%%%%%%%%%%%%%%%%%%%%%%%%%%%%%%%%%%%%%%%%%%%%%%%%

Proximity-induced superconductivity \cite{Proximity} has gained a great deal of interest in modern condensed matter physics recently \cite{ExoticSuper} due to providing feature to generate exotic superconducting correlations in non-superconducting materials \cite{yazdani}. Indeed, the underlying lattice structure of non-superconducting materials which is intimately related to the energy spectrum of materials plays an essential role. One of the most important characteristics of energy spectrum to capture possible correlated states is to provide the most available states at  Fermi energy in which charge carriers may become unstable and for instance, tend to the formation of superconductivity \cite{SuperDensity} at low temperatures.

The combination of crystal symmetry and internal degrees of freedom in condensed-matter systems results in electronic multi-band structures with unique features. In such systems band crossings occurred at high-symmetry points or on high-symmetry lines have provided a new ground to realize free unconventional quasi-particles \cite{BanCrosHigherSpin} without having high-energy counterparts. Moreover, it has been investigated that multi-band systems can host Cooper pair with higher total angular momentum \cite{SuperHigherSpin}. In particular, band crossings including flat bands exhibit more exotic pairing symmetries  \cite{intriSupFlatExoti,intriSupFlatTopo1} and topological properties \cite{intriSupFlatTopo1,intriSupFlatTopo2}.

Dice lattice model \cite{dice1}, known also as $\mathcal{T}_3$ lattice \cite{dice2}, consists of three different sublattices (A, B and C) where the A and C sites are connected indirectly through the B sites [top-right panel of Fig. \ref{fig1}]. The effective low-energy theory of band structure of dice lattice is characterized by linear dispersive bands touching at the corners of the Brillouin zone and additional middle perfectly flat band at zero energy implying triple-band crossings  [bottom-right panel of Fig. \ref{fig1}]. Dice lattice can be realized naturally by three adjacent (111) layers of cubic lattice, such as the transition-metal oxide SrTiO$_3$/SrIrO$_3$/SrTiO$_3$ trilayer heterostructure \cite{hetrost,diceChern} or artificially in cold atomic systems \cite{optlat1,optlat2}. Note that thin films of perovskite SrIrO$_3$ have been reported to exhibit metallic behavior \cite{peSrIrO3}. Dice geometry has also been implemented by Josephson junction arrays \cite{diceJJ} which is investigated in the presence of magnetic frustration \cite{diceMF}.  

On the other hand, the presence of flat band, resulting in localization of single particle states, makes a diverging peak in the density of states close to the Fermi level which usually leads to various correlated quantum phases such as Ferromagnetic \cite{FlatFero,FlatFero1}, and particularly, pair condensate phases  \cite{FlatFero1,FlantSuper0,FlantSuper1,FlantSuper}. In the latter case, the superconducting pairing potential in flat bands is proportional to pairing interaction strength \cite{FlantSuper0}, unlike the exponentially small ones in dispersive band, signaling an essential enhancement of transition temperature. The correlated phases of bosons and fermions in the presence of magnetic flux \cite{diceInter0,diceInter} have been studied for perfectly flat band systems. Especially, it has been shown that nonzero Chern number provides a lower bound for superfluid weight in the topological flat band \cite{intriSupFlatTopo2}. However, finite superfluid weight for a flat band with zero Chern number has geometric origin \cite{intriSupFlatQuMet}. Moreover, recent studies show that intrinsic superconductivity in systems with even partially flat bands can promote superconductivity \cite{intriSupParFlatEnhac1,intriSupParFlatEnhac2} with unconventional pairing potential \cite{intriSupParFlatExoti}. In most of the previous models whose band structures contain flat bands, intrinsic mechanisms for the formation of propagating Cooper pairs with either low-energy effective theory \cite{intriSupFlatExoti,intriSupFlatTopo1,FlantSuper0,FlantSuper1,intriSupParFlatEnhac1} or lattice model \cite{FlatFero1,FlantSuper,diceInter,intriSupFlatTopo2,intriSupFlatQuMet,intriSupParFlatEnhac2,intriSupParFlatExoti} have been investigated. While there is still a little attention to the penetration of superconducting correlations from external superconductors into localized normal states of dice lattice flat band.

In this work, we explore the possibility of establishing extrinsic superconducting state in the dice lattice via proximity effect and determine its properties.
We consider superconducting/normal/superconducting (SNS) Josephson junction on dice lattice with an s-wave superconducting pairing symmetry. 
Using self-consistent tight-binding Bogoliubov-de Gennes (TBdG) formalism, we reveal proximity effect of superconductivity in the normal region and its inverse effect for principal orientations of dice lattice and uncover the role of flat band of the spectra. We find that there is a critical value for chemical potential of the S region above which for undoped N region considerable superconductivity can be induced to the N region, particularly, near the junction interface. Oppositely, below the critical value by decreasing the superconducting chemical potential and approaching zero, the pairing potential of the S regions increases while proximity-induced superconductivity in the N region decreases and becomes almost independent of the N region chemical potential. Also, supercurrent is calculated as functions of phase difference across the junction and doping levels of both S and N regions. We show, interestingly, that the supercurrent is significant (small) if the chemical potential in superconductors is smaller (larger) than the critical value, where the induced pairing magnitude is small (large), which increases with the increment of the N region chemical potential.  

%%%%%%%%%%%%%%%%%%%%%%%%%%%%%%%%%%%%%%%%%%%%%%%%%%%%%%%%%%%%%%%%%%%%%%%%%%%
\section {Model and Theory}\label{s2}
\begin{figure}[t]
\includegraphics[width=8cm]{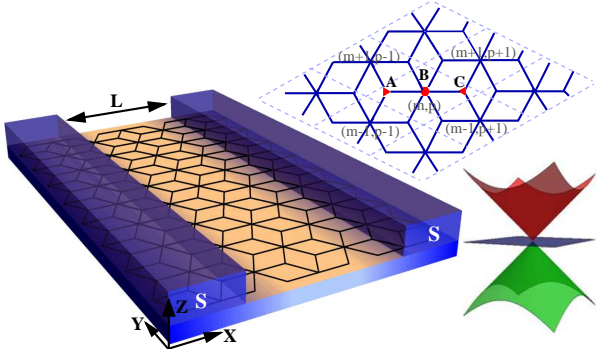}
\caption{(Color online) Left panel: Geometrical schematic of SNS junction comprising of two s-wave superconducting electrodes  deposited on top of a 2D dice lattice. $L$ is the length of N region between two superconducting leads. Zigzag direction is along the y direction whereas the system is assumed to have finite length in the x direction. Upper right panel: dice lattice structure with A, B and C are sublattices and dashed lines denote the span of unit cells with indicated their indices. Lower right panel: low-energy portion of dice lattice band structure.}
\label{fig1}
\end{figure}

We consider a dice lattice in the xy plane, as shown in Fig. \ref{fig1} (left panel), which is in contact with two superconducting leads. The two superconductors are separated by a length L and, for concreteness, assumed to have s-wave pairing symmetry. The established superconductivity on regions under the leads causes that superconducting pairing correlations would penetrate into the N region of dice lattice via proximity effect. The total Hamiltonian of itinerant carriers in the dice lattice Josephson junction can be modeled within a standard nearest-neighbor tight-binding description as,

%%%%%%%%%%%%%%%%%%%%%%%%%%%%%%%%%%%%%%%%%%%%%%%%%%%%%%%%%%%%%%%%%%%%%%%%%%%
\begin{align}
\mathcal{H}&=\mathcal{H}_{kin}+\mathcal{H}_{\Delta}\nonumber\\
\mathcal{H}_{kin}&=-t\sum_{\langle i,j\rangle ,\sigma}( c^{\dagger}_{i 1 \sigma}c_{i 2 \sigma}+c^{\dagger}_{i 3\sigma}c_{i 2 \sigma}) {+H.c}\nonumber\\
&-\sum_{\alpha=1}^3\sum_{i,\sigma} \mu(i) c^{\dagger}_{i \alpha \sigma}c^{}_{i \alpha \sigma},\\
\mathcal{H}_{\Delta}&=-\sum_{\alpha=1}^3\sum_{i,\sigma} U(i) c^{\dagger}_{i \alpha \uparrow}c^{\dagger}_{i \alpha \downarrow}c_{i \alpha \downarrow }c_{i \alpha \uparrow}+H.c,\nonumber
\label{eq1}
\end{align}
where $c^{\dagger}_{i \alpha \sigma}(c_{i \alpha \sigma})$ is electron creation (annihilation) operator of sublattice $\alpha$ with spin $\sigma=\uparrow, \downarrow$ at i{\it th} unit cell. $t$ is the nearest-neighbor hoping energy of electrons and the parameter $U(i)$ is the on-site attractive pairing interaction. $\mu(i)$ is the chemical potential having different values in the S and N regions. Experimentally, for our proposed scheme, the chemical potential of each region can be controlled individually by applying back gate voltage located at the bottom of the regions. 

Using mean-field approximation, one can decouple $\mathcal{H}_{\Delta}$ and recast it in the form
\begin{equation}
\mathcal{H}^{MF}_{\Delta}=-\sum_{\alpha=1}^3\sum_{i,\sigma} \Delta(i) c^{\dagger}_{i \alpha \uparrow}c^{\dagger}_{i \alpha \downarrow}+H.c,
\label{eq2}
\end{equation}
where the mean field superconducting order parameter $\Delta(i)$ is defined as
\begin{equation}
\Delta(i)=- \frac{U(i)}{3}\sum_{\alpha=1}^3 \langle c_{i \alpha \downarrow}c_{i \alpha \uparrow}\rangle.
\label{gapEq}
\end{equation}

Following the approach used for bipartite lattices, {\it e.g.,} graphene, in Refs. \cite{BlackS,BlackS2}, one can extend its method to three sublattices case. Thus, the total Hamiltonian can be diagonalized by using the TBdG formalism through the Bogoliubov-Valatin transformation \cite{Bogoliubov-Valatin,BlackS,JacobL},
\begin{align}\label{Bogoliubov-Valatin}
\left(
\begin{array}{c}
  c_{i1\sigma}\\
   c_{i2\sigma}\\
   c_{i3\sigma}
    \end{array}
 \right)&=\sum_{\nu=1}^{3n}
 \left(
 \begin{array}{c}
   u_{i}^{\nu}  \\
   y_{i}^{\nu}  \\
   x_{i}^{\nu}
 \end{array}
 \right)\gamma_{\nu \sigma}-
sign(\sigma)\left(
 \begin{array}{c}
    v_{i}^{\nu*} \\
    z_{i}^{\nu*}\\
   w_{i}^{\nu*}
 \end{array}
 \right)\gamma_{\nu \bar{\sigma}}^{\dagger},
\end{align}
where $\{u_{\nu}, y_{\nu}, x_{\nu}\}$ are electron and  $\{v_{\nu}, z_{\nu}, w_{\nu}\}$ are hole states in TBdG formalism. $\gamma_{\nu \uparrow}^{\dagger}(\gamma_{\nu \downarrow})$ is quasi-particle  creation (annihilation) operator in $\nu$ state with spin direction $\uparrow (\downarrow)$. The eigenvalues and eigenvectors of system can be determined by solving the following TBdG equations,\\
\begin{align}
\sum_{i}&\mathcal{H}^{MF}(i,j) \psi^{\nu}(i)= E^{\nu}\psi^{\nu}( {j}),
\label{TBdGeq}
\end{align}
with
\begin{align}
\mathcal{H}^{MF}(i,j) = \mathcal{H}_0 {(i,j)} \sigma_z +\Delta(i) \sigma_{+}+\Delta^{\dagger}(i)\sigma_{-},
\end{align}
where $\psi^{\nu}(i)=[u_i^{\nu},y_i^{\nu},x_i^{\nu},v_i^{\nu},z_i^{\nu},w_i^{\nu}]^T$ is eigenvector corresponding to eigenvalue $E^{\nu}$, $\sigma_{\pm}=(\sigma_x\pm i\sigma_y)/2$ and $\sigma_{xyz}$ are Pauli's matrices. $\mathcal{H}_0(i,j)$ and $\Delta(i)$ are $3\times3$ matrices that are defined, respectively, as:
\begin{align}\label{eq5}
\mathcal{H}_0(i,j) &= \mu(i) S_0 -t \sum_{\xi= {\pm}}\delta_{i+\xi ,j}S_{\xi},\\
\Delta(i)&=\Delta_U(i) S_0 {.}
\end{align}
Here $S_0$ is $3\times3$ identity matrix and $S_{\pm}=(S_x \pm i S_y)/\sqrt{2}$ with $S_{\nu} (\nu={x,y})$ being angular momentum matrices for spin 1 states which are given by\\
\\
$\hspace{0.5cm} S_x=\frac{1}{\sqrt{2}}\left(
   \begin{array}{ccc}
     0 & 1 & 0 \\
     1 & 0 & 1 \\
     0 & 1 & 0 \\
   \end{array}
 \right),
 $ \hspace{0.25cm} $S_y=\frac{i}{\sqrt{2}} \left(
   \begin{array}{ccc}
     0 & -1 & 0 \\
     1 & 0 & -1 \\
     0 & 1 & 0 \\
   \end{array}
 \right).$
\\
\\
Inserting Eq. (\ref{Bogoliubov-Valatin}) into Eq. (\ref{gapEq}), one can arrive at the self-consistent order parameter as
\begin{equation}
\Delta_U(i)=\frac{U(i)}{3}\sum_{\nu=1}^{3n}(u_i^{\nu}v_i^{\nu*}+y_i^{\nu}z_i^{\nu*}+x_i^{\nu}w_i^{\nu*}) \tanh(\frac{\beta E^\nu}{2}),
\label{gapEign}
\end{equation}
where $\beta =1/k_B T$ with $k_B$ is Boltzmann constant and $T$ represents temperature. So by plugging the obtained eigenvalues and eigenvectors from Eq. (\ref{TBdGeq}) into Eq. (\ref{gapEign}), the superconducting order parameter can be determined by iteration. In order to study how superconductivity penetrates from the S region into the N region, it is convenient to define pairing amplitude (PA) \cite{BlackS} as:
\begin{equation}
F_U(i)=-\Delta_U(i)/U(i).
\label{eq7}
\end{equation}
Note that the leakage of Cooper pairs from superconductors into N region, displayed by $F_U$, plays a key role in investigating the reduced as well as induced superconductivity via extrinsic mechanisms.

In what follows, we will assume that the order parameters of superconducting leads have different phases. So the supercurrent passing through the N region can be controlled by the phase gradient over the junction via the self-consistent process. Using the Heisenberg equation of motion we can obtain an expression to study supercurrent given by\\
\begin{align}\label{current}
I(i)=-e \langle\frac{dn_i}{dt}\rangle &=-e \frac{\textit{i}}{\hbar}\langle\left[ \mathcal{H}^{MF},n_i\right]\rangle \nonumber\\
&= 2 e t(\langle c^{\dagger}_{1(i-1)}c_{2 i}\rangle+\langle c^{\dagger}_{2(i-1)} c_{3 i}\rangle-H.c),
\end{align}
where $n_i$ is the particle density at i{\it th} unit cell. Usinig the Bogoliubov-Valatin transformation Eq. (\ref{Bogoliubov-Valatin}), the supercurrent (\ref{current}), which implicitly depends on phase difference between superconducting leads, can be expressed as  
\begin{align}\label{eq8}
I(i)&= 4 e t  \sum_{\nu} \lbrace (Im[x^{\nu *}_{i}y^{\nu}_{i-1}+u^{\nu}_{i-1}y^{\nu *}_i]F(E^{\nu})\nonumber\\
&+Im[w^{\nu}_{i}z^{\nu*}_{i-1}+v^{\nu *}_{i-1}z^{\nu}_i]F(-E^{\nu})\rbrace,
\end{align}
where $F(E)$ is Fermi-Dirac distribution function. Furthermore, to gain insight into the behavior of energy levels %\cite{MVHosseini} 
we investigate local density of states (LDOS) 
\begin{equation}\label{eq9}
LDOS_i(E)=\sum_{\nu}\vert \psi^{\nu}(i)\vert \delta(E-E^{\nu}),
\end{equation}
and density of states (DOS)
\begin{equation}\label{DOS}
DOS(E) = \sum_i LDOS_i(E),
\end{equation}
for all occupied states of the system.

In the following, without loss of generality, the geometry of zigzag dice nanoribbon is adopted by imposing open boundary conditions along the x-axis such that the SNS junction is oriented along the finite length of lattice with $n$ unit cells (see left panel of Fig. \ref{fig1}). We also apply Fourier transformation and periodic boundary conditions along the junction interface, which is taken to be in the y-direction. Note that though the following results are presented for the zigzag interface but for the other direction we performed the calculation and qualitatively the same results were obtained.
%%%%%%%%%%%%%%%%%%%%%%%%%%%%%%%%%%%%%%%%%%%%%%%%%%%%%%%%%%%%%%%%%%%%%%%%%%%
\section {Results and discussion} \label{s3}

Calculating superconducting order parameter self-consistently, one can obtain PA throughout the junction to find that how the existence of zero energy flat band of dice lattice affects on penetrating Cooper pairs from S to N regions.
\begin{figure}[t]
\includegraphics[width=7cm]{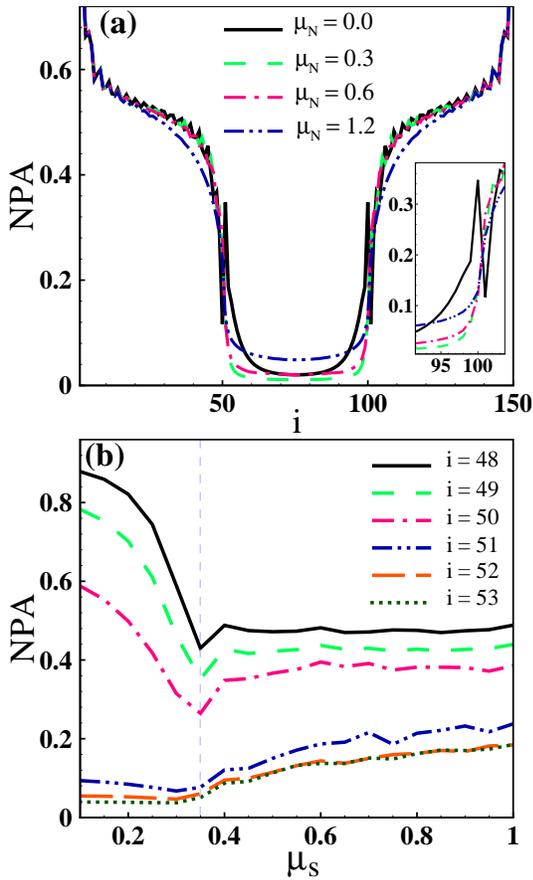}
\caption{(Color online) (a) NPA vs. lattice unit cells for different doping levels of N region with $\mu_S=1.2$, $L=50$ and 50 unit cells for S regions. (b) NPA in term of doping levels of the S regions for different unit cells near the SN interface on the S side i=48, 49, 50 and on the N side i=51, 52, 53  with $\mu_N=0.9$.}
\label{fig2}
\end{figure}
Throughout the paper U is set to be 1.36 (zero) in the S (N) region of junction, otherwise specified. Also, the temperature is chosen to be T=10 K. Moreover, the lattice constant is unity and $t$ is taken as energy unit.  
As shown in Fig. \ref{fig2}(a), the normalized PA (NPA) is plotted in terms of unit cell index for different values of doping level in the N region with superconducting doping $\mu_S=1.2$. At zero doping level $\mu_N = 0$, interestingly, there is an abrupt change of PA about the interface exhibiting a large peak (dip) on the N (S) side (see inset of Fig. \ref{fig2}(a)) with the smallest value in the middle of N region. This is can be attributed to the existence of an infinitely large DOS at zero energy level of N region causing a huge leak of Cooper pairs from the S side into the N side of the junction. As a result, giant proximity-induced superconductivity \cite{GiantProximityEffect} takes place near the interface, contrary to the flat-less band structure counterparts such as graphene Josephson junction \cite{GraphJoseph}. Furthermore, by increasing the $\mu_N$, the PA declines arising from the decrease in DOS for the N region. This is due to transition of Fermi level from the flat band to the conic band. With a further increase in $\mu_N$ once again the values of PA increase with almost uniform magnitude throughout the N region owing to approaching van Hove singularity level of the N region. Consequently, the possibility of Copper pairings in the vicinity of the interface at zero doping $\mu_N = 0$ is more than other levels.
\begin{figure}[t]
\includegraphics[width=8.7cm]{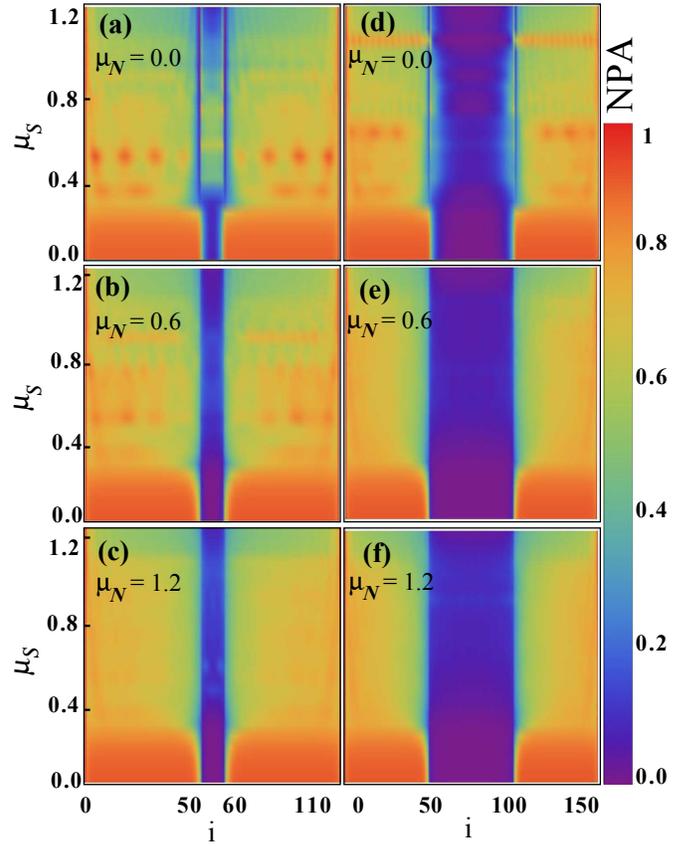}
\caption{(Color online) {NPA} along the junction as function of $\mu_S$ in three different doping levels of N region as; first row: no-doping ($\mu_N=0$), second row: moderated doping ($\mu_N=0.6$) and third row: {high doping} ($\mu_N=1.2$) for short L=10 (left column) and long L=50 (right column) junctions.}
\label{fig3}
\end{figure}

The position of the Fermi level seems to be significant in both regions. So, the dependence of PA on $\mu_S$ for some unit cells around the SN interface including both S (i $\leq$ 50) and N (51$\leq$ i $\leq$ 100) regions with doped N region is shown in Fig. \ref{fig2} (b). Importantly, one finds that there is a critical value $\mu^{\ast}_S = 0.35t$ for the chemical potential of S region, indicated by vertical dashed line, below which the PA decreases from relatively high values on the S side, similar to the superconducting surface state in rhombohedral graphite \cite{intriSupParFlatEnhac1}, and takes almost small constant values on the N side. While over the critical value the PA slightly increases and then saturates in the S region and at the same time, in the N region, it increases slowly. Remarkably, the above-mentioned behavior for the S region is in contrast to bulk superconductivity of conventional materials where by decreasing $\mu_S$ the pairing potential decreases and eventually vanishes at $\mu_S = 0$, except for high enough values of U \cite{CritUGraph}. As a result, although the order parameter has the highest value for the lowest value of $\mu_S$ establishing strong superconductivity Cooper-pair leaking distance, {\it i.e.}, direct and inverse proximity effect, becomes small (see also Fig. \ref{fig3}). This feature is very different from that of an ordinary SN interface \cite{IndCorr} and will be further discussed below.

To get a full view of the behavior, the density plots of NPA as functions of lattice sites and doping levels of the S regions are represented in Fig. \ref{fig3} for different values of $\mu_N$ and for short [Fig. \ref{fig3} (a)-(c)] and long [Fig. \ref{fig3} (d)-(f)] junctions. One observes that for S chemical potential smaller than the critical one the proximity effect and its inverse effect are not only small but also unaffected by changing the $\mu_N$. Besides of critical value of $\mu_S$, which is already discussed above, there are some visible oscillations in the S region above the $\mu^{\ast}_S$ depending on the junction length $L$ and $\mu_N$. For $\mu_N=0$ the short [Fig. \ref{fig3} (a)] and long [Fig. \ref{fig3} (d)] junctions exhibit the oscillations. But for $\mu_N=0.6$ the oscillations in the long junction [Fig. \ref{fig3} (e)] are rather spoiled out compered to the short junction [Fig. \ref{fig3} (b)]. The oscillations become further damped for both short and long junctions in $\mu_N=1.2$ [Fig. \ref{fig3} (c) and Fig. \ref{fig3} (f)]. This behavior can arise as a consequence of the combination of finite size effects and the Fermi level difference between both regions.  

\begin{figure}
\includegraphics[width=6.9cm]{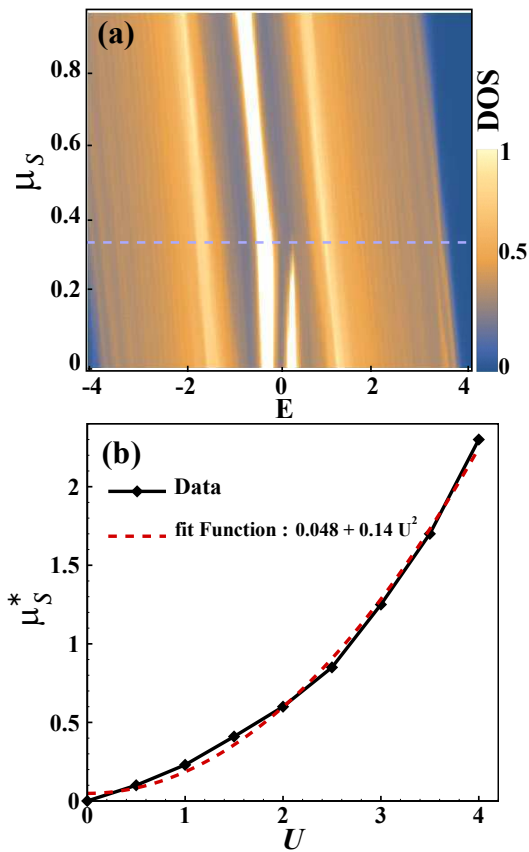}
\caption{(Color online) (a) DOS as a function of S region potential. Below $\mu_S^{\ast}$, indicated by horizontal dashed line, the superconducting gap splits the flat band developing singular DOSs at both edges of the gap. (b) Dependence of critical chemical potential $\mu_S^{\ast}$ on the pairing interaction amplitude.}
\label{fig4}
\end{figure}

In order to trace back the origin of critical chemical potential $\mu^{\ast}_S$, we studied LDOS of the S regions giving rise almost uniform states throughout the S region. Thus, the DOS of S region which can demonstrate available states as functions of $E$ and $\mu_S$ is represented in Fig. \ref{fig4} (a). One finds that as the S chemical potential is located near or at the flat band, the superconducting gap in the Fermi level can split the flat band into two flat bands making an extremely sharp coherent peak at the superconducting gap edges. Moreover, if the magnitude of doping levels is larger than $|\mu^{\ast}_S|$, the flat band is far away from the superconducting gap quenching its contribution to superconductivity. Instead, in such situation, only the dispersive bands with finite Fermi surface would participate in Cooper pairings leading to a weak pairing potential.   

Inspection above requires that the remaining parameter, namely pairing interaction U may be responsible for the values of the critical chemical potential $\mu^{\ast}_S$. Due to dependence of gap width on the pairing interaction U the range of energies, specifying  $\mu^{\ast}_S$, for which the flat band lies inside the gap is associated with U. We calculated the dependence of $\mu^{\ast}_S$ on pairing interaction $U$ as shown in Fig. \ref{fig4} (b). $\mu^{\ast}_S$ increases from zero quadratically with the paring interaction U so that strong superconductivity can be revealed by even a small U.     

\begin{figure}[t]
\includegraphics[width=6.8cm]{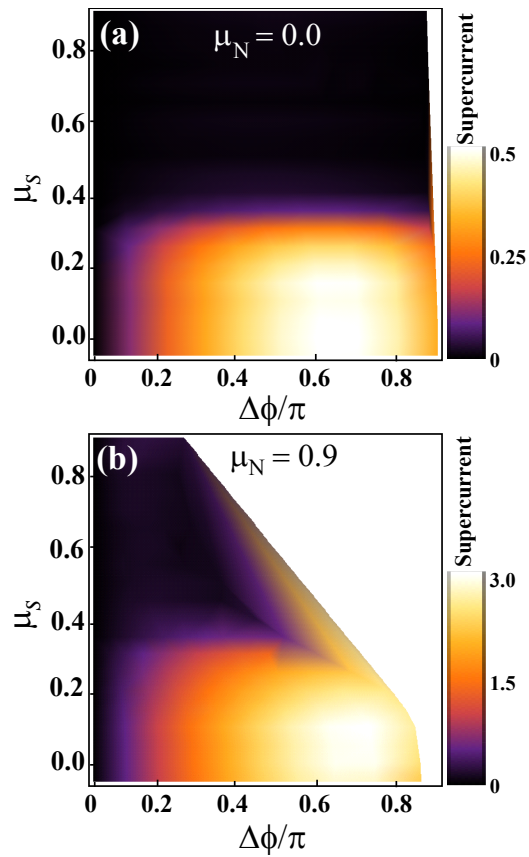}
\caption{(Color online) Supercurrent as functions of phase difference and the S region chemical potential for (a) $\mu_N = 0$ and (b) $\mu_N = 0.9$ with $L = 50$.}
\label{fig5}
\end{figure}
\begin{figure}[t]
\includegraphics[width=6.6cm]{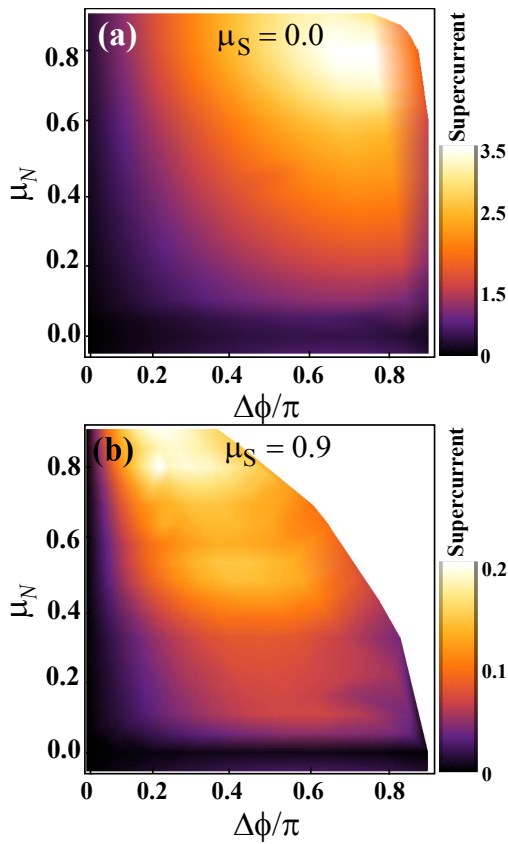}
\caption{(Color online) Supercurrent as functions of phase difference and the N region chemical potential for (a) $\mu_S = 0$ and (b) $\mu_S = 0.9$ with $L = 50$.}
\label{fig6}
\end{figure}

Correspondingly, one may expect that the Josephson current manifests spectacular behaviors due to the existence of flat band in the band structure of S and N regions. For studying the flat band effects on the supercurrent, we put the phase $\phi$ (zero) to the outermost of left (right) superconducting lead and calculate phase difference $\Delta\phi$ across the N region self-consistently as well as supercurrent. In Fig. \ref{fig5}, the supercurrent as functions of phase difference $\Delta\phi$ and doping levels in the S region is plotted with $\mu_N =0$ [panel (a)] and $\mu_N =0.9$ [panel (b)]. In both panels, the supercurrent has a maximum value below the $\mu^{\ast}_S =0.35$ implying that in the range of $\mu_S$ where the induce PA in the N region is small the pairing induced supercurrent is large and vice versa. This result can be understood as a consequence of establishing strong superconductivity below the $\mu^{\ast}_S$ in the S regions providing much more bound states carrying the suppercurrent through the junction. In contrast, for $\mu_N =0.9$ the range of supercurrent has larger value [Fig. \ref{fig5} (b)] than that of $\mu_N =0$ case [Fig. \ref{fig5} (a)]. Therefore, the supercurrent as well as superconductivity can be controlled electrically by using a small gate voltage \cite{ElectricContr}.
\begin{figure}[t]
\includegraphics[width=6.82cm]{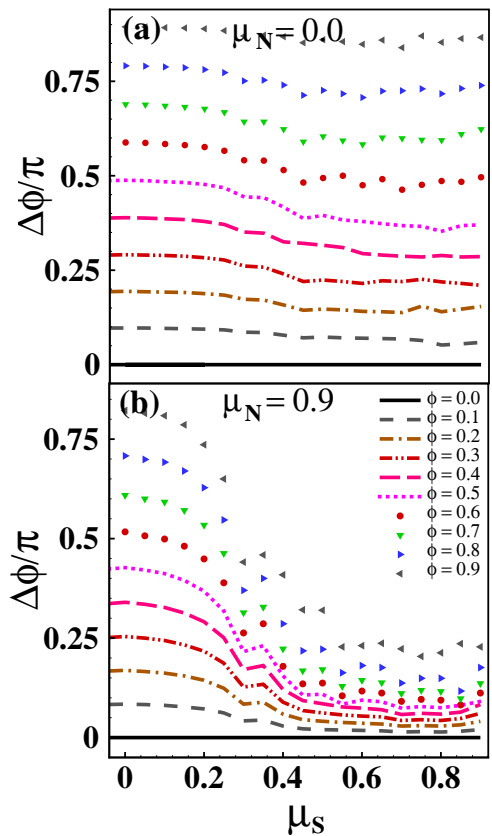}
\caption{(Color online) Phase difference as a function of the S region chemical potential for (a) $\mu_N = 0$ and (b) $\mu_N = 0.9$ with $L = 50$.}
\label{fig7}
\end{figure}

On the other hand, the dependence of supercurrent on the phase difference $\Delta\phi$ and on the chemical potential of the N region with $\mu_S =0$ and $\mu_S =0.9$ is depicted in Fig. \ref{fig6} (a) and Fig. \ref{fig6} (b), respectively. The supercurrent increases from zero by increasing $\mu_N$. Obviously, when the $\mu_N$ increases the available quasi-particle states which carry the suppercurrent increase. However, the maximum of supercurrent for $\mu_S =0$  takes place in a large phase difference but for $\mu_S =0.9$, the supercurrent is overall small with a maximum in smaller phase differences. Moreover, from both Figs. \ref{fig5} and \ref{fig6} it can be noticed that the phase difference across the N region cannot reach to large values provided both $\mu_N$ and $\mu_S$ have large values which is consistent with the previous studies \cite{BlackS}.

The evolution of $\Delta\phi$ with respect to $\mu_S$ for different $\phi$ with $\mu_N=0$ and $\mu_N=0.9$ is plotted, respectively, in Fig. \ref{fig7} (a) and Fig. \ref{fig7} (b). Below the critical chemical potential, the $\Delta\phi$ remains almost unchanged as a function of $\mu_S$ except for near the $\mu^{\ast}_S$.  Whereas above the $\mu^{\ast}_S$ the phase difference values decrease slightly (strongly) for $\mu_N=0$ ($\mu_N=0.9$) as shown in Fig. \ref{fig7} (a) (\ref{fig7} (b)). These behaviors indicate that the absence of supercurrent for $\mu_S >\mu^{\ast}_S$ (see Fig. \ref{fig6}) is mainly due to the lack of bound states rather than the small phase difference.

%%%%%%%%%%%%%%%%%%%%%%%%%%%%%%%%%%%%%%%%%%%%%%%%%%%%%%%%%%%%%%%%%%%%%%%%%%%
\section {Summary} \label{s4}

In this work, we investigated the superconducting proximity effect in heterostructure SNS involving an s-wave superconductor and a non-superconducting material based on dice lattice model containing flat band in their band structures. Using self-consistent TBdG approach, we found that there is a critical value for superconducting chemical potential depending on pairing interaction strength and determining a crossover between considerable and vanishingly small proximity effect. Below the critical chemical potential because of coinciding the Fermi level on the flat band in the S region and subsequently, splitting the flat band both direct and inverse proximity effect decrease insensitive to the N region Fermi level. When the S region chemical potential is greater than the critical value the flat band lies far away from the superconducting gap so that only conic band hosts Cooper pairs resulting in weak superconductivity. In this case, for undoped N region there is a large leak of Cooper pairs from superconductors into the N region, exhibiting a large peak near the interface owing to very large DOS in the N region. Moreover, for doped N region the proximity effect at first decreases and then increases as the doping level raises. However, by increasing $\mu_N$, the supercurrent increases from zero. But with increasing $\mu_S$, the supercurrent decreases from considerable value till the superconductor chemical potential reaches to the critical chemical potential and for $\mu_S >\mu^{\ast}_S$, it reduces to vanishingly small values. This allows for controlling the system features such as direct and inverse proximity-induced superconductivity and supercurrent simply using electric fields. 

\section*{Acknowledgment}
We would like to thank J. Linder for useful comments on this work.

%%%%%%%%%%%%%%%%%%%%%%%%%%%%%%%%%%%%%%%%%%%%%%%%%%%%%%%%%%%%%%%%%%%%%%%%%%%

\end{document}